\documentclass[iopart,twocolumn,superscriptaddress,showpacs,floatfix]{revtex4-1}
\usepackage{graphicx,amssymb}
\usepackage[fleqn]{amsmath}
\usepackage{amsfonts}
\usepackage{dcolumn}
\usepackage{bm}
\usepackage{braket}
\usepackage{multirow}
\usepackage{MnSymbol}
\usepackage{color}
\usepackage{hyperref}

\begin{document}

\date{\today}

\title{Quantifying alpha clustering in light nuclei from binding energies}

\author{K. Fossez}
\affiliation{FRIB Laboratory,
Michigan State University, East Lansing, Michigan 48824, USA}
\affiliation{Physics Division, Argonne National Laboratory, Lemont, Illinois 60439, USA}

\begin{abstract}

	What is the origin of nuclear clustering and how does it emerge from the nuclear interaction?
	While there is ample experimental evidence for this phenomenon,
	its theoretical characterization directly from nucleons as degrees of freedom remains a challenge, 
	making it difficult to improve nuclear forces using clustering observables. 
	%
	In this work, a simple ratio of energies based on effective scale arguments 
	is proposed to assess the quality of nuclear forces on alpha clustering (${ {}^{4}\text{He} }$) in bound and narrow ground-state nuclei. 
	The proposed clustering ratio is parameter-free and correctly identify basic alpha clustering features in light nuclei.  
	%
	It is found that in nuclei ranging from ${ {}^{6}\text{Li} }$ to ${ {}^{14}\text{C} }$,
	state-of-the-art \textit{ab initio} nuclear forces underestimate the degree of alpha clustering in key nuclei such as ${ {}^{8}\text{Be} }$ and ${ {}^{12}\text{C} }$.
	Stringent constraints on binding energies are then provided
	by back-propagating 10\% relative uncertainties on the experimental clustering ratios 
	using a parallel Markov chain Monte Carlo algorithm.
	It is demonstrated that the binding energies of the nuclei $^{6,7}$Li, $^7$Be, $^{10,11}$B, and $^{11}$C need to be obtained within tens of keV precision 
	by nuclear forces to ensure a proper reproduction of basic alpha clustering features in light nuclei.
	%
	%
	This study provides a new and practical path to guide future optimizations of nuclear forces with a potential impact for medium-mass nuclei.
\end{abstract}

\maketitle


\section{Introduction}

Nuclear clustering \cite{wheeler37_2315,weizsacker38_2316} refers to an emergent phenomenon by which the atomic nucleus effectively acquires a molecular-like structure and dynamics
due to the formation of tightly bound clusters of protons and neutrons at low energy \cite{oertzen06_1017,freer07_1018,beck10_b148,beck12_b149,beck14_b150,lovas13_2349,freer18_2138}.
This phenomenon is fairly common and particularly salient in light nuclei close to cluster emission thresholds \cite{okolowicz12_998,okolowicz13_241},
as summarized in the famous Ikeda diagrams \cite{ikeda68_2329}.

Light clusters are an important ingredient for the equation of state of nuclear matter \cite{hempel11_2341}  
and the symmetry energy at low temperatures and densities \cite{horowitz06_2344,natowitz10_2342,typel10_2343}, 
effectively impacting supernovae explosions and pasta phases in the crust of neutrons stars \cite{schuetrumpf17}. 

A proper understanding of nuclear clustering is still incomplete and calls for a closer link between experiment and theory \cite{freer18_2138}. 
Fundamental approaches based on nucleons as degrees of freedom play a special role in this enterprise 
since they offer the possibility to compute the degree of clustering of a $A$-body system 
defined by the probability ${ {| \braket{ A | P(A) } |}^{2} }$ where ${ P(A) }$ is a partition of the system. 

Those approaches, given a model of nuclear forces, can theoretically solve the many-body problem exactly.
In practice, only numerically controlled solutions can be provided due to finite computational resources, generating systematic errors.
They include \textit{ab initio} approaches and, it can be argued, density functional theory.

So far, no systematic calculations of clustering observables have been carried out using fundamental approaches due to their computational cost and complexity. 
However, in light nuclei, energy spectra are routinely calculated with high precision 
using modern many-body forces adjusted on two- and three-body data. 
These forces are expected to give satisfactory results in light nuclei, while in medium-mass and heavier systems it is unclear whether or not it should be the case.

Can some information about clustering be extracted from binding energies to assess the quality of nuclear forces at low cost? 
From an effective scale perspective, the emergence of clusters indicates an underlying effective separation of scales, 
which can be expressed as ratio of, for example, energies, momenta, or radii, 
associated with the transition from nucleons as degrees of freedom to nucleons and clusters as degrees of freedom.  
Providing this clustering separation of scales can be identified, 
showing its presence in a given nucleus should be enough to reveal the presence of clustering. 

Of course, the complete characterization of clustering requires more than binding energies, 
but at least, getting the associated scale separation correctly should be a prerequisite for any modern nuclear forces. 


There are deep reasons to believe that alpha (${ {}^{4}\text{He} }$) clustering specifically is an important feature of the nuclear interaction.
Nuclear matter is near a phase transition 
between a nuclear liquid and a Bose-condensed gas of alpha particles \cite{girod13_2347,ebran14_2348,elhatisari16_2311,satarov20_2339}, 
largely controlled by the strength of the interaction and its locality \cite{elhatisari16_2311}. 
A similar conclusion was reached in an earlier work \cite{ebran12_1016} connecting clustering with the depth of the nuclear mean-field,
which itself depends on the strength of the interaction.

From a different perspective, it was also shown that the Wigner SU(4) symmetric part of the interaction, 
which controls the ground state of the alpha particle, 
dominates in large nuclear systems \cite{lu19_2325}.
Moreover, it was demonstrated that adjusting a simplistic interaction on alpha-alpha scattering \cite{elhatisari15_2324} improved binding energies beyond light nuclei.

Thus, it could be argued that nuclear forces must capture alpha clustering properly to be valid beyond light nuclei, 
where the mean-field dominates and its symmetries and geometry affect properties of clustered states \cite{hecht77_2345,nazarewicz92_2346}. 

In this work, a simple ratio of energies based on effective scale arguments is proposed to quantify indirectly the degree of alpha clustering.
Using this ratio, the quality of various state-of-the-art \textit{ab initio} results based on modern nuclear forces is assessed against alpha clustering in light nuclei.
Finally, a Markov chain Monte Carlo algorithm is used to provide constraints on binding energies by back-propagating 10\% relative uncertainties on the experimental clustering ratios.


\section{From binding energies to alpha clustering}

To identify the separation of scales behind alpha clustering and express it as a ratio of energies, 
one must tell when it is energetically at least as likely to extract an alpha cluster from a $A$-body nucleus 
than to extract any four nucleons from the remaining $A-4$ nucleons. 
Alpha-emitters must be clustered by definition and will not be considered here unless the decay width is small.

The energy to extract an alpha from a $A$-body system in its ground state is the $Q_{\alpha}$-value defined as ${ {Q}_{\alpha}(A) = E(A) - E(A-\alpha) - E(\alpha) }$, 
and the energy to extract any four nucleons from the remaining $A-4$ nucleons 
is simply four times the binding energy per nucleon of the $(A-4)$-body system: ${ 4 ( E(A) - E(\alpha) )/(A-4) }$. 
This gives the condition:

\begin{equation}
	\frac{A-4}{4} \left| \frac{ {Q}_{\alpha}(A) }{ E(A) - E(\alpha) } \right| < 1 \quad (A>4),
	\label{eq_scale}
\end{equation}
which is not a general measure of the clustering scale separation, 
but should provide a decent indication of whether or not alpha clustering is present in a given nucleus.
One reminds that this condition only makes sense for experimental energies or energies obtained from many-body methods with nucleons as degrees of freedom.

In fact, one can show using perturbation theory for two weakly interacting clusters $(A-\alpha)$ and $\alpha$ (see Appendix \ref{sec_PT}) 
that at the first order the degree of one-alpha clustering, 
measured by the probability ${ {p}_{\alpha}(A) = {| \braket{ A | (A-\alpha) + \alpha } |}^{2} }$, 
is roughly proportional to ${ 1/{Q}_{\alpha}(A) + 1/{Q}_{\alpha}^{2}(A) }$ for small ${Q}_{\alpha}$-values. 
This is to say that the degree of alpha clustering should increase as the ${Q}_{\alpha}$-value decreases as one would expect, 
and this is consistent with satisfying the condition in Eq.~\eqref{eq_scale}. 

Instead of calculating the scale separation ratio in Eq.~\eqref{eq_scale}, which simply goes to zero the alpha cluster decouples from the remaining $(A-4)$ nucleons, 
the following one-alpha clustering ratio is used:

\begin{equation}
	{r}_{1}(A, \alpha) = \left| \frac{ E(A) - E(\alpha) }{ {Q}_{\alpha}(A) } \right| = \left| 1 + \frac{ E(A-\alpha) }{ {Q}_{\alpha}(A)} \right|.
	\label{eq_r1}
\end{equation}
It is the inverse of Eq.~\eqref{eq_scale} without the mass factor, 
and it diverges when ${ {Q}_{\alpha}(A) }$ tends to zero, making it sensitive near complete factorization. 
The goal is to have a simple quantity tied to alpha clustering in some conditions, and sensitive enough to reveal possible discrepancies between theory and experiment. 

The clustering ratio ${ {r}_{1} }$ is model-independent in the sense that there are no parameters 
and different many-body approaches using the same interaction should provide the same energy spectra and hence the same clustering ratios.

Long-lived excited states can also be approximately considered by removing their excitation energy ${ {E}^{*}(A) }$ to the numerator and denominator.

One notes that the magnitude of the clustering ratios is not meaningful in itself, only differences between nearby nuclei and between theory and experiment are. 
Results for the one-alpha clustering ratios using experimental energies are given in Tab.~\ref{tab1} for all nuclei where an estimate can be provided.


\begin{table}[h]
	\caption{One-alpha clustering ratios calculated using experimental data \cite{ensdf,tunl}.}
	\begin{ruledtabular}
		\begin{tabular}{clcll}
			Nucleus		& ${ {r}_{1}(A,\alpha) }$	& Nucleus		& ${ {r}_{1}(A,\alpha) }$	\\
			\hline \\[-6pt]
			$^5$He		&	0.927		&	$^{11}$Be	&	4.471	\\
			$^6$He		&	1.000		&	$^{12}$Be	&	4.505	\\
			$^7$He		&	1.057		&	$^8$B			&	5.476	\\
			$^8$He		&	1.000		&	$^9$B			&	16.58	\\
			$^5$Li		&	1.166		&	$^{10}$B	&	8.171	\\
			$^6$Li		&	2.503		&	$^{11}$B	&	5.529	\\
			$^7$Li		&	4.439		&	$^{12}$B	&	5.127	\\
			$^8$Li		&	2.885		&	$^{10}$C	&	6.278	\\
			$^6$Be		&	1.007		&	$^{11}$C	&	5.983	\\
			$^7$Be		&	5.860		&	$^{12}$C	&	8.668	\\
			$^8$Be		&	313.3		&	$^{13}$C	&	6.461	\\
			$^9$Be		&	12.10		&	$^{14}$C	&	6.411	\\
			$^{10}$Be	&	4.948		& 					&	
		\end{tabular}
	\end{ruledtabular}
	\label{tab1}
\end{table}

As expected, the helium isotopes do not exhibit any particular pattern since the excess neutrons around the alpha core are not clustered.
In the other isotopic chains, the ratio ${ {r}_{1} }$ is maximal for ${ {}^{7}\text{Li} }$, ${ {}^{8}\text{Be} }$, ${ {}^{9}\text{B} }$, and ${ {}^{12}\text{C} }$.
These results are consistent with what is already known about the structure of light nuclei and can serve as a reliable basis to test nuclear models.

One notes that using only ${ 1/|{Q}_{\alpha}(A)| }$ instead of ${ {r}_{1}(A) }$, 
and thus not accounting for how strongly bound are the remaining $A-4$ nucleons, 
would have led to ${ {}^{6}\text{Li} }$, ${ {}^{8}\text{Be} }$, ${ {}^{8,9}\text{B} }$ equally, and ${ {}^{10}\text{C} }$ 
as the most alpha clustered nuclei in their respective isotopic chains. 

The Hoyle state at 7.65 MeV above the ground state of ${ {}^{12}\text{C} }$, 
yields to a one-alpha clustering ratio of ${ {r}_{1} = 193.8 }$.
%
This large value is consistent with the characteristic clustered nature of the Hoyle state 
as supported by \textit{ab initio} calculations \cite{chernykh07_2327,epelbaum11_938,epelbaum12_2318,carlson15_1610}. 

For comparison, the nearby $3^-$ state at 9.64 MeV gives $r_1$ = 23.78 indicating some clustering but not as much as in the Hoyle state. 

In Ref.~\cite{elhatisari17_2328}, it was suggested that the ${ {0}_{2}^{+} }$ state of $^{14}$C at 6.59 MeV could have a similar structure than the Hoyle state.
This hypothesis is supported by the fairly large clustering ratio $ r_1 = 12.97 $ of this state, but the magnitude of clustering seems quite reduced in comparison. 

Binding energies given by many-body approaches can be used as well to compute the clustering ratios. 
Indeed, in light nuclei ($A \leq 12$), many-body calculations can be fully converged and clustering correlations properly accounted for. 
Consequently, if discrepancies are observed between calculated and experimental clustering ratios, it must come from the nuclear forces employed. 

In this work, both DFT and \textit{ab initio} results are used. 
The former are based on energy density functionals adjusted over thousands of ground state energies across the nuclear chart.
They are not designed to give accurate results in light nuclei but are efficient at capturing bulked properties of nuclei such as deformation.
The latter are based on modern nuclear forces usually adjusted on two- and three-body data only, and should be precise in the light sector,
but they can miss emergent properties of nuclei if those are not fully encoded into the interaction. 

There are, of course, other well-known methods based on nucleons as degrees of freedom that could not be used in this work but should be mentioned,
such as the antisymmetrized molecular dynamics \cite{ono92_2334,enyo01_2335} and the fermionic molecular dynamics \cite{feldmeier90_2336,feldmeier00_2337},
both being particularly well suited for studying clustering and with a long history summarized in Ref.~\cite{freer18_2138}. 

A fair comparison between methods using the clustering ratios requires systematic calculations based on the same Hamiltonian and at the same level of convergence. 
Unfortunately, none could be found for these methods. 

\begin{figure}[h]
	\includegraphics[width=1.0\linewidth]{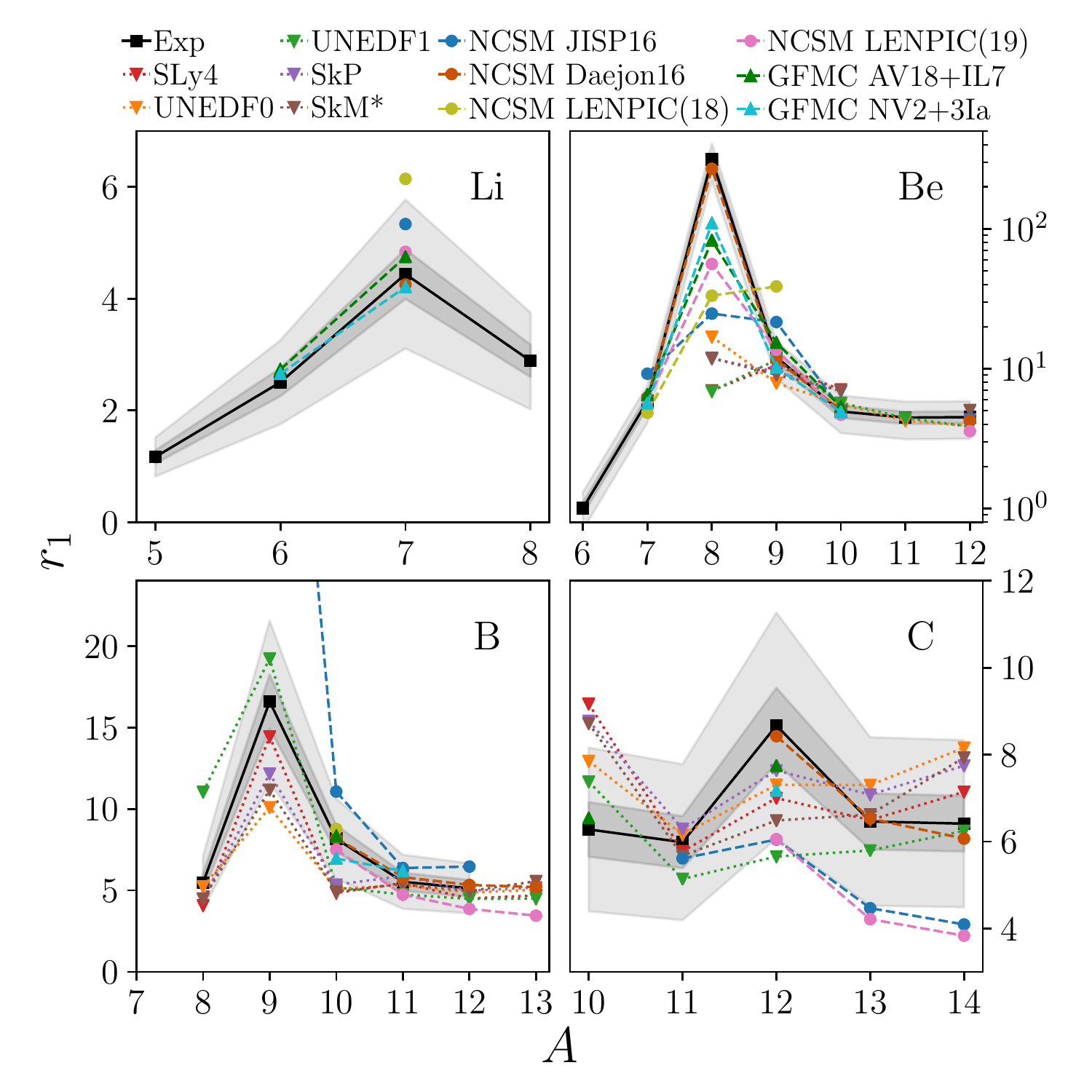}
	\caption{One-alpha clustering ratios for different DFT and \textit{ab initio} calculations (ground-state energies) compared with experimental results in the Li, Be, B, and C isotopic chains. The light gray and gray bands correspond to the 30\% and 10\% relative error bands on the experimental results, respectively.}
	\label{fig1}
\end{figure}

In Fig.~\ref{fig1},
the DFT results from Refs.~\cite{erler12_1297,massexplorer}, denoted SLy4, UNEDF0, UNEDF1, SkP, and SkM$^*$,
were used to compute the one-alpha clustering ratios of several Li, Be, B, and C isotopes and compared to experiment \cite{ensdf,tunl} (see Tab.~\ref{tab1}).
Also shown are the \textit{ab initio} Green's function Monte Carlo (GFMC) results
based on the so-called AV18+IL7 high-precision two- and three-body potentials compiled in Ref.~\cite{carlson15_1610},
as well as the GFMC results in Ref.~\cite{piarulli18_2132} based on the local chiral two- and three-body potentials including intermediate $\Delta$-excitations and denoted NV2+3Ia;
and the no-core shell model (NCSM) results based on the JISP16 \cite{maris13_2332} and Daejon16 \cite{shirokov16_2331,arxivPM19NTSE18,privcom} two-body interactions,
as well as the LENPIC interactions of 2018 \cite{binder18_2302} (two-body) and 2019 \cite{epelbaum19_2303,maris18_2333} (two- and three-body).

In some \textit{ab initio} approaches, the occasional absence of results in very light nuclei ($A \leq 5$)
was compensated using experimental data assuming they would be described correctly anyway.

First, DFT models do capture alpha clustering to a large extent but, like most other approaches presented here,
they tend to underestimate it except in ${ {}^{10,14}\text{C} }$ where they overestimate it.
Those models are efficient at describing bulk properties of nuclei and so they represent the minimum that more precise approaches should be able to achieve.

Second, the \textit{ab initio} approaches based on two-body forces only,
\textit{i.e.} the NCSM results using the JISP16, Daejon16, and LENPIC(18) interactions,
give very different results depending on how they have been optimized.

The LENPIC(18) interaction is fitted only on two-body data and gives mixed results, 
with alpha clustering overestimated in ${ {}^{7}\text{Li} }$ and ${ {}^{9}\text{Be} }$ but vastly underestimated in ${ {}^{8}\text{Be} }$.

The JISP16 and Daejon16 interactions were fitted on few-body data as well as several ground-state energies in larger systems with ${ A \leq 16 }$,
including ${ {}^{6}\text{Li} }$ and ${ {}^{10}\text{B} }$ (see next section).
However, the former did not benefit from modern fitting techniques at the time it was designed and lacks a bit of precision, 
which is why the JISP16 results do not match experiment.
This outcome can seem puzzling because this interaction can describe ${ (n,\alpha) }$ and ${ (p,\alpha) }$ scattering data \cite{shirokov18_2330},
as well as alpha-alpha scattering as shown in Ref.~\cite{kravvaris19_2326} using a new \textit{ab initio} method tailored for clustering.

The Daejon16 interaction, on the other hand, gives a near-perfect match to experiment.
This interaction, unlike JISP16, is based on chiral effective field theory potentials 
but should be able to describe the same scattering data because of the way it was built. 
The only real difference is the quality of its optimization on binding energies.
It is thus possible to build a pure two-body interaction compatible with the one-alpha clustering ratio 
as well as ${ (n,\alpha) }$, ${ (p,\alpha) }$, and alpha-alpha scattering data,
providing that some information beyond few-body systems is included in the optimization.

Finally, the \textit{ab initio} results including two- and three-body forces adjusted on few-body data only,
\textit{i.e.} the NCSM-LENPIC(19) and both GFMC results, give the correct trends and are consistent with each other 
even if they underestimate clustering significantly in ${ {}^{8}\text{B} }$ and ${ {}^{12}\text{C} }$ (outside the 10\% relative uncertainty band).
The comparison between the LENPIC interactions of 2018 and 2019 in $^9\text{Be}$ provides clear evidence of the crucial role played by three-body forces when fitting only few-body data, 
despite an underperformance of the latter in ${ {}^{13,14}\text{C} }$.

Overall, the results in Fig.~\ref{fig1} present the first evidence of a systematic deficiency in most modern nuclear forces with respect to alpha clustering.
On the one hand, \textit{ab initio} approaches based on two-body forces give satisfactory results solely when adjusted on binding energies of $A \leq 16$ nuclei,
at odd with the \textit{ab initio} rationale. 
On the other hand, those including three-body forces and adjusted on $A \leq 3$ data show consistency but underestimate alpha clustering in critical nuclei.

From a strict \textit{ab initio} perspective,
either current forces are not constrained enough to encode clustering effects,
or, as was suggested in Ref.~\cite{freer18_2138}, those effects come at higher orders in chiral effective field theory,
or the nuclear interaction should be redesigned as to explicitly account for the existence of clusters in nuclear matter.

\section{From alpha clustering to binding energies}

The experimental one-alpha clustering ratios ${ {r}_{1}^{\text{exp}}(A) }$ can be used to provide some guidance 
for the development of new nuclear forces better accounting for alpha clustering. 
Indeed, reproducing these ratios should be almost equivalent of having the right effective separation of scale behind alpha clustering, 
which is the most basic condition to satisfy. 

A realistic goal would be to reproduce the ratios ${ {r}_{1}^{\text{exp}}(A) }$ within 10\% relative uncertainties, 
as shown by the dark gray band in Fig.~\ref{fig1}. 
In fact, most \textit{ab initio} results are already close to this goal.

To reach the goal set above, one has to quantify the precision needed on each binding energy 
so that the calculated ratios ${ r_1(A) }$ fall into the 10\% relative uncertainty band around the ratios ${ {r}_{1}^{\text{exp}}(A) }$. 

This is a standard Bayesian inference problem where uncertainties on a fixed output must be back-propagated on the input
so that the latter is compatible with the former.
In this work, this is done using a Markov chain Monte Carlo (MCMC) algorithm
based on the Metropolis-Hastings method as commonly used in different fields of physics.

The problem is analogous to a network calculation in nuclear astrophysics \cite{mumpower17_2340}, albeit computationally much less expensive.
To compute one set of ${ {r}_{1}(A) }$ values, all the binding energies of all the nuclei in Tab.~\ref{tab1}, as well as those of their partition subsystems, are required.
Those energies are initialized randomly around the corresponding experimental values ${ \pm 30\% }$ to start relatively far from the ideal values.
Then they are updated by the MCMC algorithm at each step as in Ref.~\cite{mumpower17_2340},
which is itself guided by whether or not the ${ {r}_{1}(A) }$ values are approaching the experimental ones within the accepted uncertainties,
as defined by the ${ {\chi}^{2} }$ measure at a step $i$:

\begin{equation}
	{\chi}^{2}(i) = \frac{1}{ {N}_{\text{dof}} } \sum_{A} \frac{ {( {r}_{1}^{\text{exp}(A)} - {r}_{1}^{(i)}(A) )}^{2} }{ {( \Delta {r}_{1}(A) )}^{2} }.
	\label{chi2}
\end{equation}

The MCMC run is stopped when the energies are stable and the ${ {\chi}^{2}(i) }$ does not decrease significantly.
The two hyper-parameters controlling these criteria are the number of MCMC steps ${ {N}_{\text{steps}} }$ and ${ \Delta E }$ controlling the "speed" at which the parameter space is explored.
In the present work, sufficient convergence was found for ${ {N}_{\text{steps}} = 5000 }$ and ${ \Delta E = 10 \, \text{keV} }$.

At the end of one MCMC run, one set of binding energies (input) compatible with the experimental one-alpha clustering ratios (output) is obtained.
The principle of parallel MCMC is to repeat this process many times to sample the distribution of the input compatible with the output,
and then to extract mean values and standard deviations of individual binding energies, assuming Gaussian distributions.

In practice, the binding energies of $^{2,3}$H and $^{3,4}$He were kept fixed to their experimental values since they are very well reproduced theoretically.
One additional exception was considered for ${ {}^{8}\text{Be} }$ to ensure the stability of the MCMC algorithm.
Its binding energy was allowed to change, but its clustering ratio was not included in the ${ {\chi}^{2} }$ because of its very large value.

The results from the averaging of 20 runs of 50 independent MCMC runs with 5000 steps each are shown in Tab.~\ref{tab3}.
In principle, a single run of 50 independent MCMC runs would be enough,
but the additional averaging procedure ensures a better precision and allows to check that the error on the mean energies and their standard deviations is small.

\begin{table}[h!]
	\caption{Mean binding energies (in MeV) and standard deviations (in keV) of light nuclei from the back-propagation of 10\% relative uncertainties on the clustering ratios $r_1$ given in Tab.~\ref{tab1} (except for ${ {}^{8}\text{Be} }$). The experimental binding energies (in MeV) are shown for reference.}
	\begin{ruledtabular}
		\begin{tabular}{cccccccc}
			Nucleus	& ${ {E}_{\text{exp}} }$	& ${ \mu(E) }$ & ${ \sigma(E) }$ & Nucleus	& ${ {E}_{\text{exp}} }$	& ${ \mu(E) }$ & ${ \sigma(E) }$ \\
			\hline \\[-6pt]
			$^5$He     &	-27.56  &	-27.44	&	114	&	 $^{12}$Be  &	-68.65	&	-68.65	&	181 \\
			$^6$He     &	-29.27  &	-29.27 	&	96	&	 $^8$B      &	-37.74	&	-37.75	&	215 \\
			$^7$He     &	-28.86  &	-28.86	&	137	&	 $^9$B      &	-56.31	&	-56.32	&	152 \\
			$^8$He     &	-31.4   &	-31.39	&	141	&	 $^{10}$B   &	-64.75	&	-64.77	&	8 	\\
			$^5$Li     &	-26.33  &	-26.33	&	143	&	 $^{11}$B   &	-76.2 	&	-76.22	&	20 	\\
			$^6$Li     &	-31.99  &	-32.00	&	3		&	 $^{12}$B   &	-79.57	&	-79.58	&	169 \\
			$^7$Li     &	-39.25  &	-39.25	&	4		&	 $^{13}$B   &	-84.45	&	-84.44	&	216 \\
			$^8$Li     &	-41.28  &	-41.28	&	136	&	 $^{14}$B   &	-85.42	&	-85.42	&	217 \\
			$^9$Li     &	-45.34  &	-45.34	&	207	&	 $^8$C      &	-24.81	&	-24.82	&	219 \\
			$^6$Be     &	-26.92  &	-26.91	&	138	&	 $^9$C      &	-39.04	&	-39.04	&	210 \\
			$^7$Be     &	-37.6   &	-37.61	&	2		&	 $^{10}$C   &	-60.32	&	-60.31	&	165 \\
			$^8$Be     &	-56.5   &	-56.49	&			&	 $^{11}$C   &	-73.44	&	-73.45	&	15 	\\
			$^9$Be     &	-58.16  &	-58.21	&	124	&	 $^{12}$C   &	-92.16	&	-92.16	&	159 \\
			$^{10}$Be  &	-64.98  &	-64.99	&	120	&	 $^{13}$C   &	-97.11	&	-97.17	&	146 \\
			$^{11}$Be  &	-65.48  &	-65.47	&	176	&	 $^{14}$C   &	-105.3	&	-105.3	&	144
		\end{tabular}
	\end{ruledtabular}
	\label{tab3}
\end{table}

The 30 binding energies are well constrained by the optimization of the 24 clustering ratios $r_1$ as indicated by the mean values ${ \mu(E) }$. 
When looking at standard deviations, nuclei fall into two groups defined by ${ \sigma(E) \approx 100-250 }$ keV and ${ \sigma(E) < 30 }$ keV.
The latter contains only the six nuclei $^{6,7}$Li, $^7$Be, $^{10,11}$B, and $^{11}$C 
whose ground-state energies are the most constrained by the one-alpha clustering ratio.

Adding one alpha to the nuclei $^2$H, $^3$H, and $^3$He give $^{6,7}$Li, and $^7$Be, respectively, and adding one more alpha give the nuclei $^{10,11}$B, and $^{11}$C. 
Since the energies of $^2$H, $^3$H, $^3$He, and the alpha were kept fixed, only the actual binding energy of $^{6,7}$Li, and $^7$Be were changing in their clustering ratios. 
This explains why their energies are more strongly constrained than in other nuclei. 
In fact, the same happens for $^8$Be and for the same reason, if its clustering ratio is included in the MCMC run, 
but it makes calculations unstable and hence more difficult to converge. 

However, this fact alone does not explain why the energy variations ${ \sigma(E) }$ are so small. 
For example, in $^7$Li, a naive evaluation of its clustering ratio alone by varying only its binding energy around the experimental value   
reveals that the energy can be lowered by about 360 keV, or raised by about 290 keV, and still give a ratio within the 10\% relative uncertainty band. 
This is larger than the 4 keV found and indicates that the constraints on $\sigma(E)$ come from the other nuclei through the minimization of the ${ \chi^2 }$. 

Knowing that the binding energies of ${ {}^{5}\text{He} }$ and ${ {}^{5}\text{Li} }$ do not exhibit a strong sensitivity to alpha clustering 
makes the situation of the JISP16 interaction less puzzling.
Also, the reason why the Daejon16 interaction matches experimental ${ {r}_{1} }$ values is presumably because it reproduces 
the binding energies of the nuclei ${ {}^{6}\text{Li} }$ and ${ {}^{10}\text{B} }$.

The case of ${ {}^{10}\text{B} }$ provides some support for a link between three-body forces and alpha clustering.
It was found in \textit{ab initio} calculations \cite{navratil07_733,mccutchan12_2321,freer18_2138}
that this nucleus is particularly sensitive to three-body forces. 
More specifically, it was shown in Ref.~\cite{nollett07_944} that the "Illinois" (IL) family of three-body forces used in the GFMC-AV18+IL7 calculations
were able to describe correctly ${ (n,\alpha) }$ scattering data,
and that the IL7 version, in particular, yielded an improved ordering of the low-lying states in $^{10}$B \cite{carlson15_1610}. 
Moreover, it is also known that ${ {}^{9}\text{Be} }$ and ${ {}^{9}\text{B} }$, which are one proton and one neutron away of ${ {}^{10}\text{B} }$, respectively,
have ground states very close to thresholds and hence are prone to clustering \cite{rodriguez10_2319,garrido10_2320},
supporting the idea that ${ {}^{10}\text{B} }$ is clustered and has a $\alpha-\alpha-d$ structure.
Finally, the results in Tab.~\ref{tab3} show that ${ {}^{10}\text{B} }$ is one of the most sensitive nuclei with respect to alpha clustering,
and the results in Fig.~\ref{fig1} show that three-body forces are critical to describe alpha clustering in light nuclei. 
These elements together should motivate future \textit{ab initio} studies in that direction.

To summarize, the MCMC procedure revealed that the binding energies of six light nuclei, 
made of the light clusters $^2$H, $^3$H, $^3$He, and $^4$He, 
are strongly constrained by alpha clustering. 
This information could be beneficial for future optimizations of \textit{ab initio} nuclear forces.


\section{Conclusions}

In conclusion, 
the empirical one-alpha clustering ratio introduced in this work, 
based solely on binding energies, allow to correctly identify
the nuclei $^7$Li, $^8$Be, $^9$B, and $^{12}$C as being the most clustered in their respective isotopic chains.  

It was shown that, according to the one-alpha clustering ratio,
state-of-the-art \textit{ab initio} approaches based on modern nuclear forces significantly underestimate alpha clustering in key nuclei
such as ${ {}^{8}\text{Be} }$ and ${ {}^{12}\text{C} }$.

Finally, relative uncertainties on the one-alpha clustering ratio were back-propagated on binding energies using a MCMC method
and revealed that the ground states energies of $^{6,7}$Li, $^7$Be, $^{10,11}$B, and $^{11}$C are particularly sensitive to alpha clustering.

This work opens new avenues ranging from the extension of the clustering measures beyond ${ {}^{14}\text{C} }$ to the study og other types of clusters. 
Perhaps, the most interesting opportunity is the development of new two- and three-body forces in the spirit of the ${ \text{NNLO}_{\text{sat}} }$ interaction \cite{ekstrom15_1766}
including the binding energies of some of the six nuclei identified in their optimization.  
Such interactions could give satisfying results in medium-mass nuclei
studied at the Facility for Rare Isotope Beams (FRIB) \cite{balantekin14_2338}.


%

In all fairness, 
the measures introduced are empirical and mostly provide \textit{a posteriori} tests of the presence of clusters. 
For these reasons, a genuine attempt was made at quantifying alpha clustering to reveal real issues in nuclear models,
but conclusions were mostly limited to qualitative statements.

A picture is emerging where alpha clustering, three-body forces, and observables in medium-mass nuclei are interrelated.
Alpha clustering is an important feature of the nuclear interaction
and should be used when developing and evaluating modern nuclear interactions 
as was recently done in Ref.~\cite{arxivKravvaris20} for three-body forces on ${ (n,\alpha) }$ scattering.

\appendix

\section{One-alpha clustering measure approximation in perturbation theory}
\label{sec_PT}

The isospin symmetric approximation of the one-alpha clustering measure ${ {r}_{1}(A, \alpha) \approx | 1 + E(A-\alpha) / {Q}_{\alpha}(A) | }$ 
can be justified approximately using perturbation theory. 
Assuming two non-interacting clusters $\alpha$ and $a$ with their non-orthogonal ground and first excited states $\ket{\alpha, a}$ and $ \ket{\alpha, a^*} $, respectively, 
the state of the interacting system $A = \alpha + a$ writes in the first order of perturbation theory:

\begin{equation}
	\ket{A} = \ket{\alpha, a} + \varepsilon \frac{ \braket{ \alpha, a^* | V | \alpha, a } }{ E(\alpha,a) - E(\alpha,a^*) } \ket{ \alpha, a^* },
	\label{eq_A_PT}
\end{equation}
where $\varepsilon$ is a small parameter and $V$ is the interaction between the two clusters. 
The interaction matrix elements will be written using the short notation $ \braket{ \alpha, a^* | V | \alpha, a } = V_{a^*,a} $ for convenience. 
The energy of the total system in its first excited state is:

\begin{equation}
	E(A^*) = E(\alpha,a^*) + \varepsilon V_{a^*,a^*}.
	\label{eq_EA_PT}
\end{equation}
Using the fact that $ E(\alpha,a) = E(\alpha) + E(a) $ (non-interacting clusters) 
and denotinig the energy difference between the ground state and the first excited state $ \delta E(A) = E(A^*) - E(A) $, 
The energy difference $ E(\alpha,a) - E(\alpha,a^*) $ can be written as:

\begin{equation}
	E(\alpha,a) - E(\alpha,a^*) = E(\alpha) + E(a) - E(A) + \delta E(A) + \varepsilon V_{a^*,a^*},
	\label{eq_Ediff_PT}
\end{equation}
where one recognizes the $Q_{\alpha}$-value. 
The overlap between the interacting and factorized systems then writes:

\begin{equation}
	\braket{\alpha, a | A} = 1 + \frac{ \varepsilon V_{a^*,a} }{ \delta E(A) - Q_{\alpha}(A) + \varepsilon V_{a^*,a^*} } \braket{ \alpha, a | \alpha, a^* }.
	\label{eq_overlap_PT}
\end{equation}
For small perturbations, \textit{i.e.} $ \varepsilon << 1 $, one obtains:

\begin{equation}
	\braket{\alpha, a | A} \approx 1 + \frac{C}{ \delta E(A) - {Q}_{\alpha}(A) },
	\label{eq_overlap_PT_approx}
\end{equation}
where $C$ is a small constant. 
The degree of clustering is then given by the probability ${ {p}_{\alpha}(A) = {| \braket{ A | \alpha,a } |}^{2} }$, 
indicating possible corrections in $1/Q_{\alpha}^2(A)$ to the one-alpha clustering measure ${ {r}_{1}(A, \alpha) }$.

\begin{acknowledgments}
I am grateful to M. Piarulli for sharing the binding energies obtained using the GFMC-NV2+3Ia interaction,
as well as to P. Maris and J. Vary for sharing the last update of the binding energies obtained using the NCSM-JISP16, Daejon16, and LENPIC(19) interactions,
and for the ensuing discussions.
I also want to thank N. Vassh and S. K\"onig for clarifying the use of the parallel MCMC algorithm and for pushing me to think beyond the one-alpha clustering ratio 
during their productive visits at Argonne National Laboratory.
My appreciation also goes to R. Wiringa, H. Hergert, S. Bogner, and W. Nazarewicz for their patience and useful comments on clustering and nuclear forces.
Finally, I would like to thank T. Li and S. Wang for pointing out errors in the original manuscript. 
This material is based upon work supported by the U.S. Department of Energy,
Office of Science, Office of Nuclear Physics, under the FRIB Theory Alliance award DE-SC0013617.
An award of computer time was provided by the Institute for Cyber-Enabled Research at Michigan State University.
\end{acknowledgments}


%

\end{document}